\documentclass{iau} 
\usepackage{graphicx}

\title[The first detections of the key prebiotic molecule PO in star-forming regions
] 
{The first detections of the key prebiotic molecule PO in star-forming regions}

\author[V\'ictor M. Rivilla]   
{V\'ictor M. Rivilla$^1$,
Francesco Fontani$^1$,
Maite Beltr\'an$^1$,
Anton Vasyunin$^2$,
Paola Caselli$^2$,
Jes\'us Mart\'in-Pintado$^3$
 \and Riccardo Cesaroni$^1$}

\affiliation{$^1$Osservatorio Astrofisico di Arcetri \\ Largo Enrico Fermi, 50125,
Florence, Italy \\ email: {\tt rivilla@arcetri.astro.it} \\[\affilskip]
$^2$Max Planck Institute for Extraterrestrial Physics, \\ Giessenbachstrasse 1, 85748, Garching, Germany  \\ [\affilskip]
$^3$ Centro de Astrobiolog\'ia (CSIC-INTA) \\ Ctra. de Torrej\'on a Ajalvir km 4, 28850, Torrej\'on de Ardoz, Spain \\ 
}

\pubyear{2017}
\volume{332}  
\jname{Astrochemistry VII – Through the Cosmos from Galaxies to Planets}
\editors{Maria Cunningham, Tom Millar and Yuri Aikawa, eds.}
\begin{document}

\maketitle

\begin{abstract} Phosphorus is a crucial element in prebiotic chemistry, especially the P$-$O bond, which is key for the formation of the backbone of the deoxyribonucleic acid. So far, PO had only been detected towards the envelope of evolved stars, and never towards star-forming regions.
We report the first detection of PO towards two massive star-forming regions, W51 e1/e2 and W3(OH), using data from the IRAM 30m telescope. PN has also been detected towards the two regions. The abundance ratio PO/PN is 1.8 and 3 for W51 and W3(OH), respectively. Our chemical model indicates that the two molecules are chemically related and are formed via gas-phase ion-molecule and neutral-neutral reactions during the cold collapse. The molecules freeze out onto grains at the end of the collapse and desorb during the warm-up phase once the temperature reaches $\sim$35 K. The observed molecular abundances of 10$^{-10}$ are predicted by the model if a relatively high initial abundance of phosphorus, 5$\times$10$^{-9}$,  is assumed.

\end{abstract}

\firstsection 
\section{Introduction}

The detection of new interstellar molecules related with prebiotic chemistry in star-forming regions will allow us to make progress on understanding how the building blocks of life could originate in the interstellar medium (ISM). However, there is a key ingredient that still evades detection:
phosphorus, P. This element is essential for life (\cite[Maci\'a et al. 1997]{macia97}), because it plays a central role in the formation of P-bearing macromolecules such as phospholipids, which are the structural components of all cellular membranes, or adenotryphosphate (ATP), responsible for the transfer of energy in cells (\cite[Pasek et al. 2005]{pasek05}). Especially important to basic biochemistry is the P$-$O bond, fundamental for many relevant biological molecules such as phosphate esters, which are essential for the formation of the backbone of the genetic macromolecule deoxyribonucleic acid (DNA).

Phosphorus is thought to be synthesized in massive stars and injected to the ISM through supernova explosions (\cite[Koo et al. 2013]{koo13}). It has a cosmic abundance of P/H$\sim$ 3$\times$10$^{-7}$ (\cite[Grevesse \& Sauval 1998]{grevesse98}). 
It has been detected towards atmospheres of stars (\cite[Caffau et al. 2011]{caffau11}, \cite[Roederer et al. 2014]{roederer14}, \cite[Caffau et al. 2016]{caffau16}), but it is barely detected in the ISM. The ion P$^+$ has been detected in several diffuse clouds (\cite[Jura \& York 1978]{jura78}), and only a few simple P-bearing species (PN, PO, CP, HCP, C$_2$P, PH$_3$) have been identified towards the circumstellar envelopes of very evolved objects (\cite[Tenenbaum et al. 2007]{tenenbaum07}, \cite[De Beck et al. 2013]{debeck13}, \cite[Ag\'undez et al. 2014]{agundez14}). In star-forming regions, only PN has been detected so far (\cite[Turner \& Bally 1987]{turner87}; \cite[Ziurys \& Friberg 1987]{ziurys87}; \cite[Fontani et al. 2016]{fontani16}). Previous searches of PO towards star-forming regions (\cite[Sutton et al. 1985]{sutton85}; \cite[Matthews et al. 1987]{matthews87}) were unsuccessful.

\begin{figure*}
\centering
\includegraphics[scale=0.45]{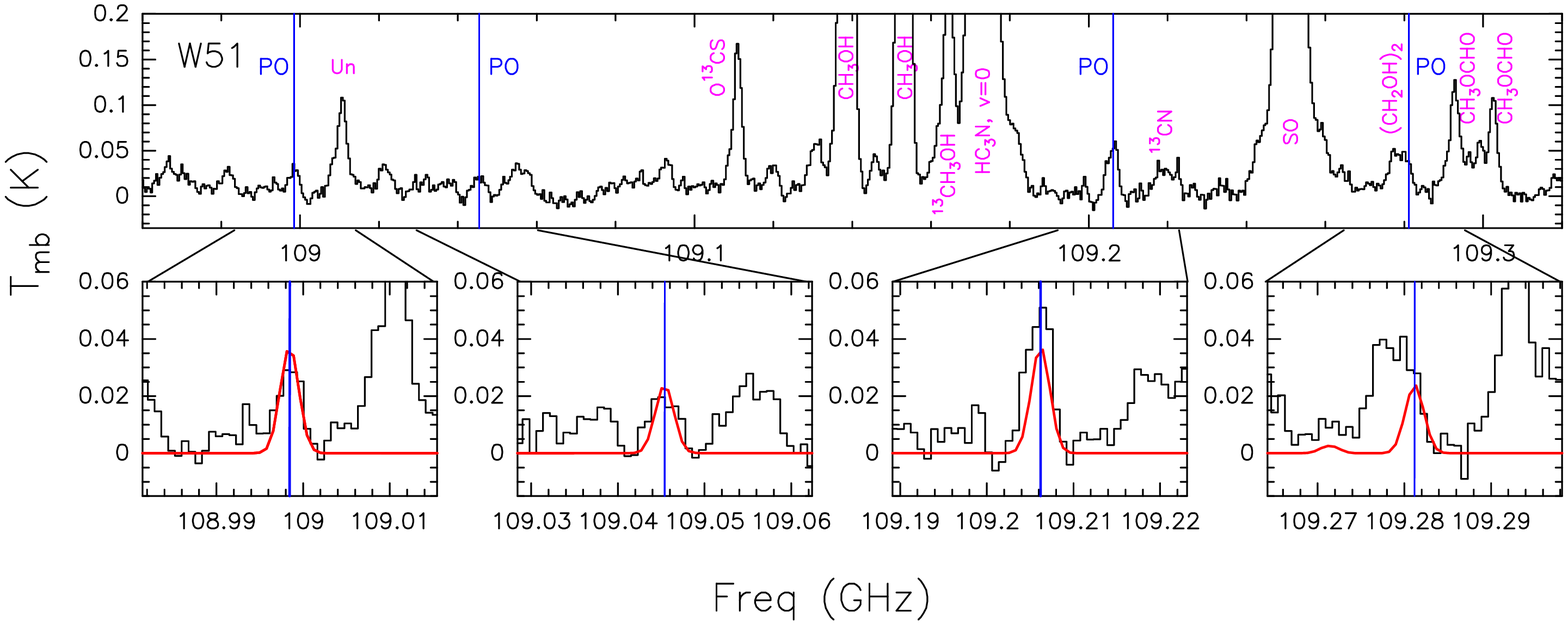}
\caption{Spectrum observed at 3 mm towards W51. The PO transitions are indicated with blue vertical lines. The lower panels show zoom-in views of the PO transitions. The red line is the LTE fit.}
\label{figure-PO-W51-3mm}
\end{figure*}

Despite the prebiotic importance of P-bearing molecules, their chemistry is still poorly understood in the ISM. 
The few theoretical models devoted to P-chemistry disagree in both the chemical formation pathways and the predictions of the abundances of PO and PN.
While some works (\cite[Millar et al. 1987]{millar87}, \cite[Adams et al. 1990]{adams90}, \cite[Charnley \& Millar 1994]{charnley94}) suggest that PN should be more abundant than PO, other studies involving theoretical modeling and laboratory experiments 
(e.g. \cite[Thorne et al. 1984]{thorne84}) predict that P should be found mainly in the form of PO.
To constrain the chemical models and understand the chemistry of phosphorus in the ISM, astronomical detections of PO and PN in star-forming regions are required. 

We present observations searching for PN and PO towards two massive star-forming regions: the W51 e1/e2 and the W3(OH) complexes (with luminosities of 1.5$\times$10$^{6}$ and 1.0$\times$10$^{5}$ L$_{\odot}$, respectively). 

\section{Observations and Results}

We used data from different observing runs performed with the IRAM 30m telescope at Pico Veleta (Spain) on 2012 and 2015. The W51 e1/e2 (hereafter W51) region was observed at 1, 2 and 3 mm, while the W3(OH) region was observed at 3 mm. The spectra were exported from the software package CLASS of GILDAS to MADCUBA (Madrid Data Cube Analysis on ImageJ is a software developed in the Center of Astrobiology, Madrid) to visualize and analyze single spectra and datacubes; see \cite[Rivilla et al. 2017a]{rivilla17a}), which was used for the line identification and analysis. 

\subsection{Detection of PN}

\begin{figure*}
\centering
\includegraphics[scale=0.4]{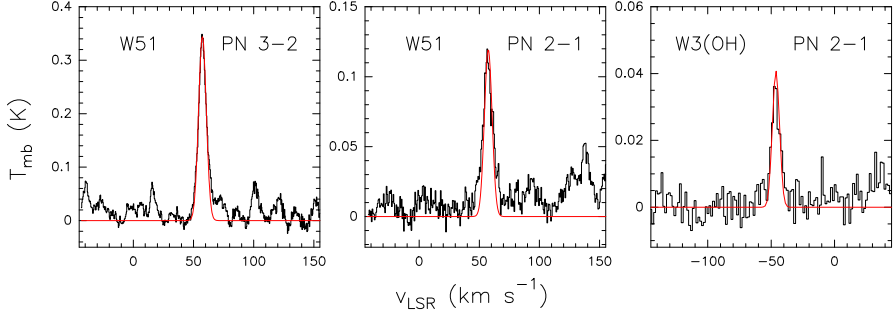}
\caption{PN transitions observed towards W51 and W3(OH). The red line is the LTE fit.}
\label{figure-PN}
\end{figure*}

We detected the $J$=2$-$1 and $J$=3$-$2 transitions of PN towards W51 (left and middle panels of Fig \ref{figure-PN}).
Using an angular diameter of $\sim$12 arcsec, the LTE simulated spectra reproduce the PN $J$=2$-$1 and $J$=3$-$2 transitions for an excitation temperature of $\sim$35 K (Fig. \ref{figure-PN}), as previously found by \cite[Turner et al. (1987)]{turner87}. We derived a PN source-averaged column density of 2.1$\times$10$^{13}$ cm$^{-2}$. 
This translates into a molecular abundance with respect to molecular hydrogen of 1.1$\times$10$^{-10}$.  

We also detected the $J$=2$-$1 PN transition towards W3(OH) (right panel of Fig. \ref{figure-PN}). Assuming the same excitation temperature and linear diameter as that of W51 (i.e. 30 arcsec at the distance of W3(OH)), we obtained a source-averaged PN column density of 0.2$\times$10$^{13}$ cm$^{-2}$, which implies a molecular abundance of 0.4$\times$10$^{-10}$.

\subsection{Detection of PO}

The PO radical is a diatomic molecule with a dipole moment of $\mu$=1.88 D (\cite[Kanata et al. 1988]{kanata88}). Its millimeter spectrum was measured in the gas phase in the laboratory by \cite[Kawaguchi et al. (1983)]{kawaguchi83}. 
PO has a $^2\Pi_{\rm r}$ ground electronic state, and therefore each $J$-transition splits into a doublet because of $\Lambda$ coupling. The spin of I$=$1/2 of the phosphorus atom produces a further splitting of each $\Lambda$ doublet into two hyperfine levels.


We used MADCUBA to identify the PO transitions in the observed spectra. We assumed the same temperature ($T_{\rm ex}$=35 K) and source size as that of PN to produce the LTE spectrum.
We show the 3 mm spectra of W51 and the LTE fit in Fig. \ref{figure-PO-W51-3mm}. We clearly identified the PO quadruplet towards W51. The four transitions are well fitted by the LTE model. The lines are detected above 6$\sigma$. These PO transitions were already detected in the wind of the oxygen-rich AGB star IK Tauri by \cite[De Beck et al. (2013)]{debeck13}. 



The LTE fit of the four PO spectral features at 3 mm gives a source-averaged PO column density of 4$\times$10$^{13}$ cm$^{-2}$ and a molecular abundance of 2$\times$10$^{-10}$. The predicted PO lines at at 1 and 2 mm are consistent with the observed spectra. 
In summary, we have detected all the 12 detectable PO transitions towards W51 with intensities consistent with LTE emission: two of them (at 3 mm) are unblended, two (at 3 mm) are partially blended, while the others (all the 2 mm and 1 mm transitions) are heavily blended.

For W3(OH), PO was also detected. We have fitted the lines with the LTE model, assuming $T_{\rm ex}$=35 K and the same diameter assumed for PN (30 arcsec), and obtained a column density of 0.6$\times$10$^{13}$ cm$^{-2}$ and a molecular abundance of 1.2$\times$10$^{-10}$.

\section{Discussion: the chemistry of phosphorus}

We investigated the formation of the phosphorus-bearing molecules PO and PN in massive star-forming regions using the model based on \cite[Vasyunin \& Herbst (2013)]{vasyunin13}, which simulates the chemical evolution of a parcel of gas and dust with time-dependent physical conditions. First, the formation of a dark dense clump from a translucent cloud during free-fall collapse is simulated (cold starless phase). During this phase, the density increases from 10$^3$ cm$^{-3}$ to 10$^{6}$ cm$^{-3}$ and visual extinction rises from A$_v$=2 to A$_v\gg$100 mag. Gas and dust temperatures are both decreasing slightly from 20 K to 10 K. On a second phase, the dense dark clump warms up from 10 K to 200 K during 2$\times$10$^{5}$ years, thus developing into a hot core (warm-up protostellar phase). We defined as time 0 the moment when the protostar starts to heat up the environment.

We used two sets of initial chemical abundances to explore the influence of the poorly constrained P depletion factor: i) the low metals model (EA1, \cite[Wakelam \& Herbst, 2008]{wakelam08}), where the atomic P is depleted and has an abundance of 2$\times$10$^{-10}$ with respect to hydrogen, typically used in dark cloud chemical models; and ii) the high metals model, where the phosphorus abundance is increased by a factor of $\sim$25 to 5$\times$10$^{-9}$.

\begin{figure*}
\centering
\includegraphics[scale=0.45, angle=0]{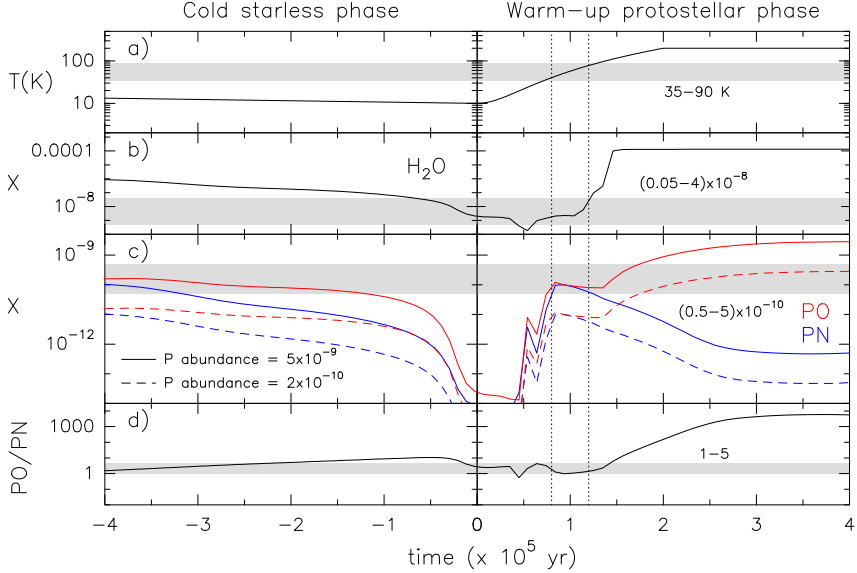}
\caption{\small Results of our chemical model. The left panels correspond to the cold collapse phase, and the right panels to the subsequent warm-up phase. We show the evolution during 0.8$\times$10$^{5}$ yr, being time 0 the moment when the protostar starts to heat up the environment. The two vertical dotted lines in the right panels indicate the temporal range for which the observational constraints are fulfilled (see text).
 {\it a)} Evolution of the temperature assumed by the model. The gray band corresponds to the temperature range between 35 and 90 K which is compatible with the observational constraints (see text).  
{\it b)} H$_2$O abundance. The gray band corresponds to the range found towards massive star-forming regions by \cite[Marseille et al. (2010)]{marseille10}.
{\it c)} PO and PN abundances predicted by the model (red and blue lines, respectively). We have considered two different initial abundances of depleted atomic phosphorus: 5$\times$10$^{-9}$ (solid lines, high metals model), and 2$\times$10$^{-10}$ (dashed lines, "low metals" model, usually assumed in dark cloud models). 
The gray band indicates a conservative range of one order of magnitude around the molecular abundances derived from our observations of PO and PN (see text).
{\it d)} Ratio of the molecular abundances of PO and PN. The gray band shows a conservative range (1$-$5) around the ratio found by the observations.}
\label{figure-model}
\end{figure*}

The results of the modeling are shown in Fig. \ref{figure-model}, which shows the temporal evolution of the temperature (Fig. \ref{figure-model}a), gas abundance of water (Fig. \ref{figure-model}b), gas abundances of PO and PN (Fig. \ref{figure-model}c) and the PO/PN ratio (Fig. \ref{figure-model}d). 
The two P-bearing species are chemically related and are formed purely in a sequence of gas-phase ion-molecule and neutral-neutral reactions. PO is efficiently formed during the cold collapse phase in a sequence of reactions:

\begin{equation}
{\rm H_3O^+ + P  \longrightarrow HPO^+ + H_2}
\end{equation}
\begin{equation}
{\rm HPO^+ + e^- \longrightarrow PH + O}
\end{equation}
\begin{equation}
{\rm HPO^+ + e^- \longrightarrow PO + H}
\end{equation}
\begin{equation}
{\rm P^+ + H_2  \longrightarrow PH_2^+}
\end{equation}
\begin{equation}
{\rm PH_2^+ + e^- \longrightarrow PH + H}
\end{equation}
\begin{equation}
{\rm O + PH \longrightarrow PO + H}
\end{equation}

PN is mainly formed from PO via:

\begin{equation}
{\rm N + PO \longrightarrow PN + O}
\end{equation}

During the starless phase the model predicts that PO is more abundant than PN (Fig. \ref{figure-model}c). Both species freeze out to grains at the end of the cold collapse phase and consequently the gas abundances sharply decrease (Fig. \ref{figure-model}c). PO has an abundance $\sim$5 times higher than PN at that time.

At the warm-up phase, once the temperature gradually increases from 10 K to 200 K (Fig. \ref{figure-model}a), the species evaporate to the gas phase in the order determined by their desorption energies. PO and PN desorb simultaneously when the temperature reaches $\sim$35 K ($\sim$0.7$\times$10$^{5}$ yr, Fig. \ref{figure-model}c), because in our model they have the same evaporation energy of 1900 K
.

Some PO is converted to PN via gas-phase reactions, and the abundances of the two species become almost equal until $\sim$1.2$\times$10$^{5}$ yr (Fig. \ref{figure-model}c). Once the temperature reaches $\sim$100 K, water ice evaporates and the gas-phase abundance of water reaches a high value of 10$^{-4}$ (see Fig. \ref{figure-model}b). Then, the abundance of protonated water H$_3$O$^+$ correspondingly increases, and PN reacts with H$_3$O$^+$:

\begin{equation}
{\rm PN + H_3O^+ \longrightarrow HPN^+ + H_2O}
\end{equation}

In turn, HPN$^+$ has two equally probable channels of dissociative recombination:

\begin{equation}
{\rm HPN^+ + e^- \longrightarrow PN + H}
\end{equation}
\begin{equation}
{\rm HPN^+ + e^- \longrightarrow PH + N}
\end{equation}

Since the abundance of atomic oxygen is higher by almost an order of magnitude than the abundance of atomic nitrogen, PH is converted to PO rather than PN. As such, PN is gradually destroyed, while PO is additionally produced, which significantly increases the PO/PN ratio (Fig. \ref{figure-model}d).

To compare the results of the observations with the predictions of our chemical models, we show the values of the abundances and the PO/PN ratio derived from the observations in Figs.~\ref{figure-model}c and \ref{figure-model}d, respectively. 
The main source of uncertainty of the estimated abundances arises from the uncertainty of the size of the emitting region and the H$_2$ column density. We have considered a conservative range of one order of magnitude uncertainty for the abundances, i.e. (0.5$-$5)$\times$10$^{-10}$ (see Fig.~\ref{figure-model}c). In the case of the PO/PN ratio, which is independent of the H$_2$ column density, we have considered a lower uncertainty of a factor of 5, i.e., values in the range 1$-$5. These observational constraints can only be explained by the "high metals" model (Fig. \ref{figure-model}c), and during a period of $\sim$5$\times$10$^{4}$ yr (between 0.8 and 1.2$\times$10$^{5}$ yr in Fig. \ref{figure-model}). This suggests that the value of the initial depleted atomic abundance of P is a factor of $\sim$25 higher (i.e. 5$\times$10$^{-9}$) than that commonly assumed by the standard "low metal" model. 

The temporal window that fulfills the observational constraints translates in a temperature range between the desorption temperature of the P-bearing molecules (35 K) and $\sim$90 K (see Fig.~\ref{figure-model}a). The former temperature is the same than that derived from PN transitions towards W51, which may indicate that the observations have detected the bulk of PN (and PO) that have recently desorbed from dust grains. The latter temperature is just below the sublimation temperature of the bulk of water (100 K), when PN is efficiently destroyed and partially converted to PO. Then, the gas abundance of water predicted by the model at this stage is still low, $\sim$10$^{-9}$, which is in good agreement with the values of 5$\times$10$^{-10}$ $-$ 4$\times$10$^{-8}$ found towards several massive star-forming regions by \cite[Marseille et al. (2010)]{marseille10} (see Fig.~\ref{figure-model}b).

\section{Conclusions}

We report the first detection of the key prebiotic molecule PO towards two star-forming regions: W51 e1/e2 and W3(OH). The derived molecular abundances of PO are $\sim$10$^{-10}$ in both sources. We have found an abundance ratio PO/PN of 1.8 and 3 for W51 e1/e2 and W3(OH), respectively, in agreement with the values estimated for evolved stars. 
Our chemical modeling indicates that the two molecules are chemically related and are formed via gas-phase ion-molecular and neutral-neutral reactions during the cold collapse. The molecules freeze out onto grains at the end of the collapse, and evaporate during the warm-up phase once the temperature reach $\sim$35 K.
Similar abundances of PO and PN are expected during a period of $\sim$5$\times$10$^4$ yr at the early stages of the warm-up phase, when the temperature is in the range 35$-$90 K. The observed molecular abundances require a relatively high initial abundance of atomic phosphorus of 5$\times$10$^{-9}$, 25 times higher than the $"$low-metal$"$ P-abundance typically used in dark cloud chemical models.

\end{document}